%% file: main.tex
\documentclass[journal]{IEEEtran}
\usepackage{xcolor}
\usepackage{balance}
\usepackage{cite}
\usepackage{amsmath,amssymb,amsfonts}
\usepackage{graphicx}
\usepackage{textcomp}
\usepackage{acronym}
\usepackage{xcolor}
\usepackage{tikz} 
\usepackage[utf8]{inputenc}
\usepackage{pgfplots} 
\usepackage{pgfgantt}
\usepackage{pdflscape}
\usepackage{changes}
\usepackage{comment}
\usepackage{subfigure}
\usepackage{tabularx}
\usepackage{acronym}
\usepackage{mathtools,algpseudocode,algorithm,MnSymbol}
\usepackage[inline]{enumitem}
\usepackage{makecell}
\usepackage{multirow}
\newcommand{\hthickline}{\noalign{\hrule height 0.80pt}}
\usepackage[table]{xcolor}
\usepackage{pgfplots}
  \pgfplotsset{compat=newest}
  \usetikzlibrary{plotmarks}
  \usetikzlibrary{arrows.meta}
  \usepgfplotslibrary{patchplots}
  \usepackage{grffile}
  \usepackage{amsmath}

\pgfplotsset{compat=newest} 
\pgfplotsset{plot coordinates/math parser=false} 

\input{acronyms}

\newcommand{\rev}[1]{{#1}}

\begin{document}

\bibliographystyle{IEEEtran}
\bstctlcite{IEEEexample:BSTcontrol}
\title{Sensing with Mobile Devices through Radio SLAM: \\ Models, Methods, Opportunities, and Challenges}

\author{Yu Ge, Ossi Kaltiokallio,  Elizaveta Rastorgueva-Foi, Musa Furkan Keskin, Hui Chen, Guillaume Jornod, \\ Jukka Talvitie, Mikko Valkama, Frank Hofmann, and Henk Wymeersch\vspace{-5mm}}

\maketitle

\begin{abstract}
The integration of sensing and communication (ISAC) is a cornerstone of 6G, enabling simultaneous environmental awareness and communication. This paper explores radio SLAM (simultaneous localization and mapping) as a key ISAC approach, using radio signals for mapping and localization. We analyze radio SLAM across different frequency bands, discussing trade-offs in coverage, resolution, and hardware requirements. We also highlight opportunities for integration with sensing, positioning, and cooperative networks. The findings pave the way for standardized solutions in 6G applications such as autonomous systems and industrial robotics.
\end{abstract}


\vspace{-4mm}
\section{Introduction}

\Ac{ISAC} is anticipated to be a core enabler of 6G \cite{procIEEE_2023_ISAC_6G}, offering services beyond communication through three sensing configurations: monostatic, bistatic, and multistatic sensing. 
In the first configuration, the transmitter and receiver are co-located. In the second case, they are at different locations. In the third type, there are multiple transmitters and receivers,  distributed in different geographical locations.
Sensing information in 6G will not only enable new applications (e.g., cooperative driving and robotic interaction) but will also improve communication performance through contextual insights and digital radio twins \cite{khan2022digital}. Despite significant advances in ISAC,  challenges remain. First, monostatic sensing requires full-duplex capabilities \cite{Barneto-WCM21} or physically separate receivers, which are costly. Bistatic and multistatic sensing between \acp{BS} holds appeal due to the known and fixed geometry of transmitters and receivers, but is incompatible with standard duplexing. Thus, the most feasible ISAC configuration involves \acp{BS} interacting with  \acp{UE}. However, this approach faces the challenge that one sensing endpoint (e.g., UE) has an \textit{unknown geometric state}, making it challenging to align sensing results within a global coordinate system.

\begin{figure*}
    \centering
    \includegraphics[width=0.8\textwidth]{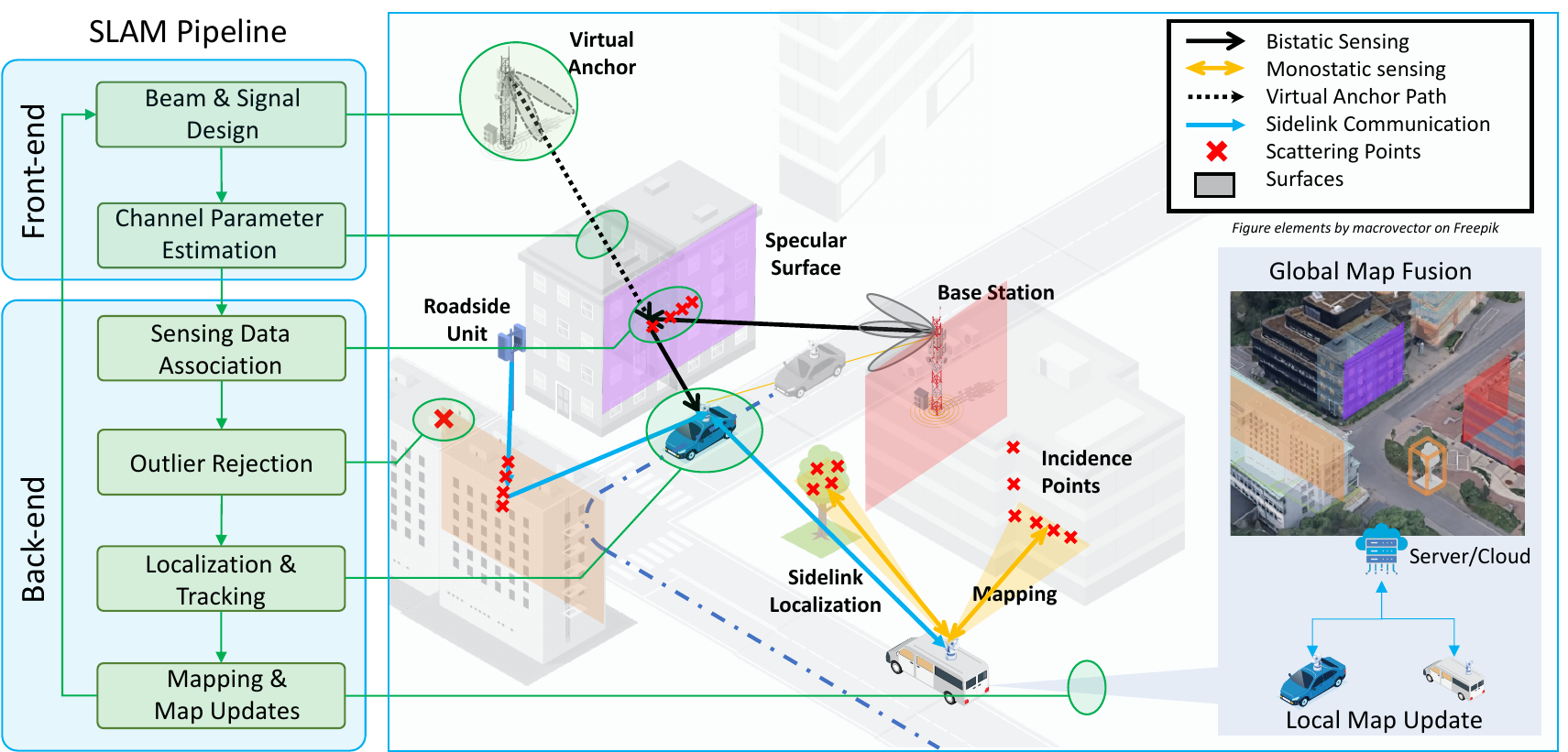}
    \caption{Key components and processes involved in a radio SLAM system, showcasing interactions between infrastructure (e.g., BSs, roadside units) and mobile devices (e.g., UE, vehicles), as well as integration with sidelink communication for enhanced data sharing and cooperative localization.  The SLAM pipeline includes critical stages like channel parameter estimation, sensing \ac{DA}, outlier rejection, and global map fusion, which enable SLAM.  }\vspace{-5mm}
    \label{fig:Radio SLAM overview}
\end{figure*}

\rev{\Ac{SLAM} addresses this challenge by linking the sensing and positioning processes \cite{durrant2006simultaneous}, thus forming a critical subroutine in bistatic ISAC with UE involvement.} \rev{{Radio SLAM} \cite{amjad2023radio}, which builds upon \ac{SLAM}'s foundations in robotics, consists of a front-end and a back-end, as depicted in  Fig.~\ref{fig:Radio SLAM overview}. The front-end leverages prior information to optimize signals, followed by channel parameter estimation to extract physical characteristics of the radio propagation paths. With appropriate data association and outlier rejection, the back-end localizes the \ac{UE} while simultaneously maintaining and updating a global map of the environment.} 
In wireless contexts, radio SLAM diverges from  robotics SLAM with two distinct characteristics: first, it leverages \rev{prior} known landmarks (i.e., BSs) that provide a fixed coordinate system, and second, it operates through bi- or multistatic measurements rather than the monostatic measurements typical in robotic SLAM (e.g., from camera, radar, or LiDAR)\rev{, providing the ability to perceive around corners}. Research on radio SLAM has therefore evolved independently, considering technologies such as UWB \cite{leitinger2019belief}, 5G \cite{witrisal2016high}, and future 6G \cite{lotti2023radio}. 
Within 5G, radio SLAM has enabled precise UE positioning with a single BS 
\cite{ge2022mmwave}. \rev{While we consider SLAM to be a subroutine of bistatic ISAC, \cite{yang2023multi} highlighted a different perspective by treating SLAM as a specific use case of ISAC, and further introduced cross-user, cross-frequency, and cross-device SLAM mechanisms. }

\rev{Radio SLAM  estimates UE trajectories and maps the environment, where key use cases} include automated vehicles, UAVs, industrial robotics, \rev{virtual and augmented reality, }and indoor positioning \rev{ and navigation}, where precise localization and mapping are crucial. 
Performance is assessed via positioning and radar sensing KPIs, including \textit{mapping accuracy} (fidelity of environmental mapping),\textit{ positioning accuracy} (UE location error), \textit{latency }(real-time updates), \textit{energy efficiency}, \textit{scalability}, and \textit{integration capability}. Standardization is essential to align use cases, models, and protocols, as explored by ITU, ETSI, and 3GPP\rev{~\cite{kaushik2024toward}}.

Now that first 6G requirements are being formulated in standardization and that a mature body of radio SLAM literature is available, it is timely to review the state of radio SLAM and explore its future within 6G ISAC. This paper provides such an overview, detailing the \textit{components}, \textit{models}, \textit{methods}, and \textit{potential for standardization} of radio SLAM, alongside a \textit{forward-looking perspective} on the opportunities and challenges radio SLAM faces in the 6G era. In addition, various numerical examples are provided, including new results compared to the state-of-the-art.

\vspace{-3mm}

\section{The Radio SLAM Front-End}

\rev{The front-end\footnote{Our definitions of front-end and back-end are  different from classical GraphSLAM, where the front-end handles raw sensor data, performs feature extraction and \acf{DA}, while the back-end performs high-level inference.} of radio SLAM focuses on extracting geometric information from the wireless channel. To interpret the wireless channel correctly, we begin by reviewing SLAM channel modeling.}

\vspace{-3mm}
\subsection{SLAM Channel Modeling}
\begin{figure*}
    \centering
\includegraphics[width=0.99\linewidth]{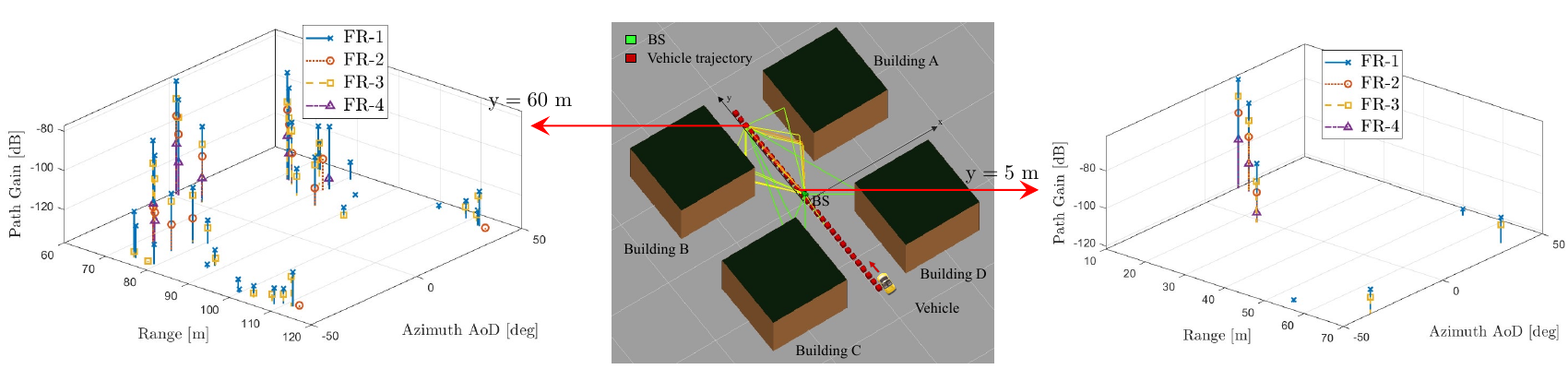}
    \caption{An urban intersection scenario with a fixed BS at $[0 \, \rm{m}, \, 0 \, \rm{m}, \, 10 \, \rm{m}]$, and a UE passing through the intersection. The left and right subfigures show channel responses across four different frequency bands-FR1 ($3.5 \,$ GHz), FR3  ($10 \,$ GHz), FR2 ($27.2 \,$ GHz) and FR4 ($140 \,$ GHz))-visualized in the delay-angle domain. These responses are obtained from ray-tracing data when the UE is at $[1.6 \, \rm{m}, \, 60 \, \rm{m}, \, 1.5 \, \rm{m}]$ and $[ 1.6\, \rm{m}, \, 5 \, \rm{m}, \, 1.5 \, \rm{m}]$, corresponding to a rich propagation condition with strong reflections from the surrounding buildings and a sparse multipath profile, respectively. The RMS delay spreads for FR1, FR3, FR2 and FR4 at $60 \, \rm{m}$ are $6.04 \, \rm{m}$, $5.80 \, \rm{m}$, $5.21 \, \rm{m}$ and $5.97 \, \rm{m}$, respectively, while those at $5 \, \rm{m}$ are $1.22 \, \rm{m}$, $1.69 \, \rm{m}$, $1.48 \, \rm{m}$ and $1.40 \, \rm{m}$.}
    \label{fig:urban_intersection}
\vspace{-4mm} \end{figure*}

\rev{A realistic channel model captures the essential characteristics of wireless channels, offering critical insights into the interaction between radio signals and the environment. Therefore, SLAM channel modeling is essential for theoretical analysis, model-based algorithm design, and performance evaluation across diverse scenarios and system configurations. In this context, the operating frequency plays a critical role in shaping SLAM performance trade-offs.} 5G and 6G frequencies, spanning FR1 (410 MHz–7.125 GHz) to FR4 (100–300 GHz) \cite{hexaxii_d43}, offer distinct SLAM trade-offs. FR1 ensures wide coverage and through-wall propagation, making it ideal for long-range, high-mobility applications, though with lower spatial resolution. FR2 (24–75 GHz) provides finer resolution for indoor and mid-range use but suffers from higher path loss. FR3 (7–15 GHz) balances coverage and resolution, while FR4 (sub-THz) offers high precision but is highly susceptible to atmospheric absorption.
Fig.~\ref{fig:urban_intersection} shows the channel responses in the delay-angle domain in four different frequency bands, obtained through ray-tracing simulations in an urban intersection scenario. Path gain decreases with increasing frequency, resulting in sparser multipath profiles at higher frequencies (FR2 and FR4) compared to lower frequencies (FR1 and FR3).\rev{\footnote{\rev{It is worth noting that a decrease in absolute path gains does not necessarily imply a smaller RMS delay spread, as seen from the values in Fig.~\ref{fig:urban_intersection} when the UE is at $60 \, \rm{m}$. This is because the RMS delay spread is determined by the \textit{normalized} path gains rather than their \textit{absolute} values. Consequently, a multipath profile with weak path gains can still produce a large delay spread if the relative delays are widely distributed.}}} In addition, when the UE (vehicle) is far from the BS, we observe a rich multipath profile due to strong reflections from the surrounding buildings, evidenced by larger \ac{RMS} delay spreads. In contrast, when the vehicle is close to the BS, the multipath profile is sparse, characterized predominantly by \ac{LoS} propagation, a significant ground reflection and other weak reflections, with correspondingly smaller RMS delay spreads.
\rev{Such} frequency-dependent channel characteristics impose \rev{different} hardware and modeling requirements for SLAM. For lower frequencies (e.g., FR1 and FR3), achieving high accuracy depends heavily on time calibration and synchronization to enhance delay domain performance. In contrast, higher frequencies, such as FR2 and FR4, require careful calibration of antenna arrays (e.g., taking into account radiation patterns and array orientation) and high-fidelity modeling of amplifiers, which are sensitive to peak-to-average power ratio (PAPR). Additionally, all bands can feature near-field effects and non-stationary characteristics, which requires careful modeling.

These considerations highlight the need for \textit{innovative solutions} to boost SLAM performance, and \textit{standardized channel and hardware models} that reflect SLAM-specific requirements.
The channel models for radio signals can be classified into stochastic and deterministic. With the inclusion of stochastic elements that account for urban, suburban, rural, and indoor scenarios with parameters like path loss, shadowing, and delay spreads, flexible and large-scale modeling can be provided for communications. However, SLAM is environment-dependent, so that  geometric and deterministic channel models are needed (e.g., ray-tracing models). These models can be customized, depending on the layout of the objects, at the expense of high computational cost. In addition, \rev{most of the} existing ray-tracing models fall short in supporting near-field effects, non-stationary channels, and the environmental interactions critical for effective SLAM operations. 

\rev{In addition to channel modeling, optimized transmit signal design constitutes a crucial component of the SLAM model as it directly impacts the received observations. The objective is to enhance channel parameter estimation and localization performance.} 
Beam design refers specifically to spatial domain optimization, involving the design of complex beamforming weights at individual antenna elements over time, while signal design can also cover time and frequency domain optimization of OFDM waveforms, such as pilot configuration and power allocation. In the presence of \textit{a priori} information about the location of the UE and landmarks, spatial design at the BS can significantly improve angle estimation accuracy while time-frequency optimizations enhance delay-Doppler estimation, although at the cost of potential grating lobes.

\vspace{-4mm}

\subsection{SLAM Channel Parameter Estimation}
\label{sec:slam_channel_parameter_estimation}

In a broad sense, the channel parameter estimation problem in radio SLAM can be defined as the task of estimating the \textit{geometric} parameters (i.e., path delays, Dopplers, \acp{AoA} and \acp{AoD}) of the propagation paths between the transmitter and the receiver in a monostatic, bistatic, or multistatic configuration. Unlike \textit{channel estimation for communications}, where the objective is to estimate the composite end-to-end channel matrix/tensor without delineating its inner \textit{geometric} structure, or \textit{channel estimation for positioning}, where the objective is to extract the \ac{LoS} parameters, 
the radio SLAM channel estimation problem involves detecting and resolving different propagation paths and accurately estimating their corresponding parameters~\cite{procIEEE_2023_ISAC_6G}. In the radio SLAM framework, this geometric information will be instrumental in localizing the connected device (i.e., UE) and static objects (i.e., landmarks or targets) to construct a complete radio environmental map \cite{rastorgueva2024millimeter}.          

The fundamental challenges of radio SLAM parameter estimation include: \textit{(i)} \textit{Complex multipath propagation:} Rich propagation environments often feature multi-bounce reflections, diffuse scattering and diffraction that complicate parameter extraction. For instance, diffuse scattering at objects results in path clusters with closely spaced angles and delays, leading to non-resolvable paths and fluctuating power levels. \textit{(ii)} \textit{Frequency band variations:} Performance may differ significantly across frequency bands. FR2 bands, with their larger bandwidths, offer fine resolution but suffer from increased path loss, while FR1 bands provide better coverage but have lower resolution and experience more pronounced edge diffraction and multi-bounce reflections, resulting in denser multipath channels.  \textit{(iii)} \textit{Stringent accuracy and latency requirements:} SLAM channel parameter estimator must deliver accurate outputs for effective localization and mapping, as minor angular inaccuracies can significantly affect location estimates, especially over long distances. This level of precision is generally unnecessary in communication systems, where angular deviations have minimal impact on performance. Moreover, these estimates must be provided within certain latency budgets, as the UE is constantly moving. 
In general, radio SLAM channel parameter estimation algorithms should be tailored to various propagation environments and frequency bands, meeting stricter accuracy and synchronization requirements than those typical of communications systems.

Various radio SLAM channel estimation solutions have been proposed in the literature, broadly categorizable into five distinct classes. First, \textit{matched filtering (MF)}-based algorithms 
apply low-complexity correlation operations to retrieve path parameters. However, they suffer from limited range and angle resolution, especially in FR1, due to low bandwidths and a small number of array elements \cite{v2x_pos_JSAC_2024}. Second, \textit{subspace}-based approaches such as \rev{\ac{ESPRIT}} and \rev{\ac{MUSIC}} can offer high-resolution estimates by exploiting specific array structures, allowing to distinguish closely spaced paths beyond Rayleigh resolution limits, but may lead to high complexity for high-dimensional channels. Third, \textit{maximum-likelihood (ML)}-based methods can provide asymptotically optimal estimates at high computational  cost, even under iterative approaches such as  \rev{\ac{SAGE}}\footnote{\rev{ Parametric channel estimation  is used in communication to reconstruct the channel for data recovery and for optimizing communication performance, whereas in SLAM, these techniques enable localization and mapping.}} and \rev{\ac{RiMAX}}. Fourth, \textit{sparsity}-based methods, such as compressive sensing and atomic norm minimization, leverage the sparse nature of channels at FR2 and above, and enable high-resolution estimation of a limited number of strong paths. However, these algorithms can be sensitive to noise and regularization terms, which can introduce additional non-zero components that cause the algorithm to misinterpret noise as part of the signal, resulting in the identification of false or spurious components. Finally, \textit{machine learning}-based methods have recently gained popularity, as they do not require any model and can naturally be integrated into the SLAM back-end.

\vspace{-5mm}
\subsection{Towards the Standardization of SLAM Front-End}

Several aspects of the radio SLAM front-end can be considered in standardization forums, including channel models, data formats, and requirements. 
Standardizing SLAM channel models is challenging due to the complexity of future wireless systems. \rev{While standardization bodies such as 3GPP are developing ISAC channel models, these are extensions of earlier stochastic models for communication and fail to capture the richness of real ISAC and SLAM channels.}
 Key factors, such as near-field effects, \rev{temporal and spatial consistency, and material properties}  must be accounted for to ensure \rev{reliable} modeling  in dynamic environments. Consistent modeling across frequency bands requires large-scale measurement data and AI-driven analysis for accurate propagation modeling.
Partial standardization (e.g., coordinate system definitions, feature models) can ensure interoperability while allowing vendor-specific innovations, especially in multi-device applications like autonomous vehicles and robotics~\cite{6g_standard_ISAC_2024}.
Additionally, defining minimum performance benchmarks is crucial for safety-critical applications such as autonomous driving and industrial automation.
\rev{Standardization} is also needed to establish protocols for data exchange and pilot signal transmission. \rev{While SLAM receiver processing can remain vendor-specific, common formats for data sharing will facilitate cooperative SLAM and sensor fusion across devices and platforms. Cooperative SLAM (coined cross-user SLAM in \cite{yang2023multi}) supports transforming, compressing, sharing, and updating of maps, allowing UEs with limited capabilities to benefit from maps constructed by high-end UEs.}

\vspace{-4mm}
\subsection{Radio SLAM Front-End  -- a Case Study}

\begin{figure}
    \centering
    \includegraphics[width=0.99\linewidth]{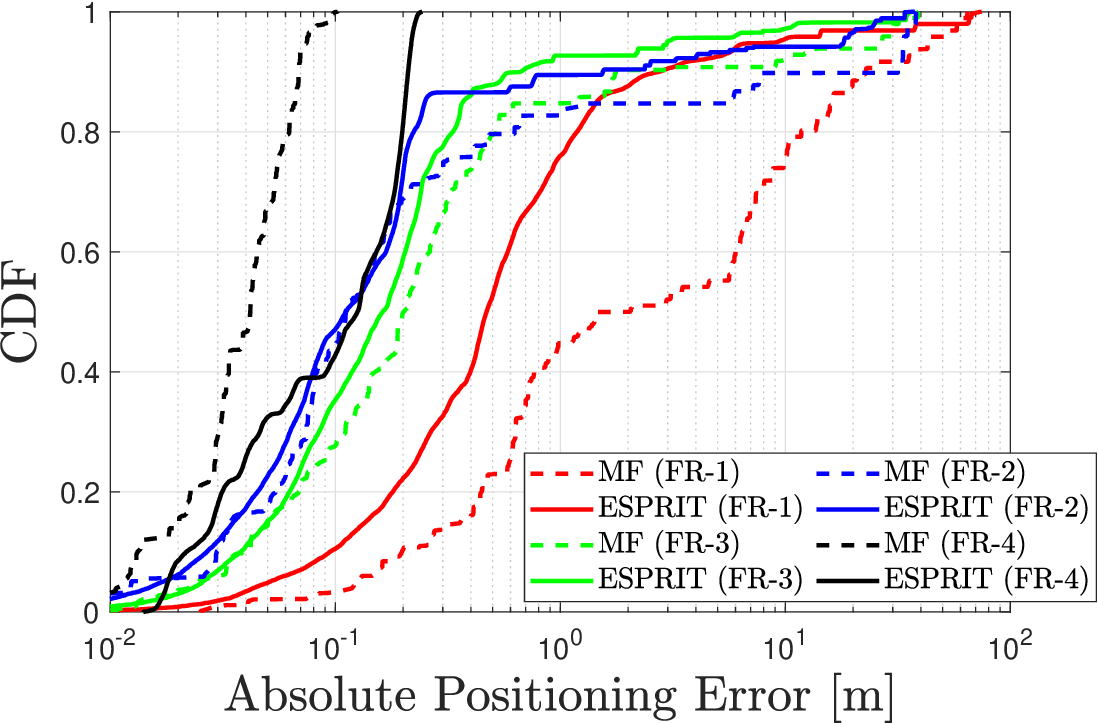}
    \vspace{-5mm}
    \caption{Empirical CDF of absolute positioning errors aggregated over $101$ different locations of a UE in an urban intersection scenario, with $100$ Monte Carlo observations generated per location. The radio SLAM scenario involves a multiple-antenna BS and a single-antenna UE communicating over a multipath channel via the round-trip time (RTT) protocol\rev{, and SLAM is performed at the BS side}. Two different SLAM channel estimation algorithms (namely, MF and ESPRIT) are employed to extract the delay and AoA/AoD of multiple paths and obtain single-snapshot position estimates of the UE. 
    Four different frequency ranges are considered, with the following parameters $\{\text{carrier frequency, bandwidth, BS array configuration}\}$. 
    FR1: $\{3.5 \, \text{GHz}, 10 \, \text{MHz}, 1 \times 2\ (4.29 \, \text{cm})\}$,
    FR3: $\{10 \, \text{GHz}, 50 \, \text{MHz}, 1 \times 4\ (4.5 \, \text{cm})\}$,
    FR2: $\{27.2 \, \text{GHz}, 100 \, \text{MHz}, 1 \times 9\ (4.41 \, \text{cm})\}$,
    FR4: $\{140 \, \text{GHz}, 1 \, \text{GHz}, 1 \times 41\ (4.29 \, \text{cm})\}$.}
    \label{fig_CE_MF_ESPRIT}\vspace{-4mm}
\end{figure}

To evaluate frequency band impact, Fig.~\ref{fig_CE_MF_ESPRIT} presents the cumulative distribution function (CDF) of \ac{UE} positioning errors\rev{, which serves as a more compact and efficient alternative to presenting separate delay and angular estimation errors,} using MF-based and ESPRIT-based algorithms \cite{v2x_pos_JSAC_2024}. The simulation, conducted in an urban intersection scenario, aggregates results from 101 UE locations using the REMCOM Wireless InSite\textregistered ray-tracer.
ESPRIT outperforms MF at FR1 and FR3, where dense multipath environments demand high path resolvability. MF struggles at these bands due to limited bandwidth and small arrays, while ESPRIT surpasses resolution limits for better accuracy. At FR2 and FR4, however, MF benefits from larger bandwidths and sparse multipath profiles, achieving accuracy comparable to or better than ESPRIT-based methods, which are inherently suboptimal. Notably, higher frequencies improve positioning accuracy despite greater path losses, as the gains in delay and angular resolution outweigh propagation challenges.

\vspace{-3mm}

\section{The Radio SLAM Back-End} 
The back-end of radio SLAM deals with tracking the  \ac{UE}, detecting and localizing landmarks, and associating measurements from the front-end to those landmarks. 

\begin{table*}[ht!]
\centering
\scriptsize
    \caption{{{A Summary of Different SLAM Algorithms}}}
    \centering
    \renewcommand{\arraystretch}{1.2}
    \begin{tabular}{| c  | c | c | c | c | c | c |}
    \hthickline
     \makecell{\textbf{Tasks}} 
    & \makecell{\textbf{Snapshot SLAM}}
    & \makecell{\textbf{EKF-SLAM}} 
    & \makecell{\textbf{FastSLAM}} & \makecell{\textbf{GraphSLAM}}
    & \makecell{\textbf{BP-SLAM}} & \makecell{\textbf{RFS-SLAM}}
    \\
    \hthickline
        \multirow{1}{*}{\rotatebox{0}{Pre-solved DA\ }} 
        & \cellcolor{blue!5}
        Yes
        & \cellcolor{blue!5}
        Yes
        & \cellcolor{blue!5}
        Yes
        & \cellcolor{blue!5}
        Yes & \cellcolor{blue!5} No & \cellcolor{blue!5} No
        \\
        \hthickline
        %
        %
        %
        %
        %
    \multirow{1}{*}{\rotatebox{0}{Correlation between map and sensor\ }} 
        & \cellcolor{red!5}
        Yes
        & \cellcolor{red!5}
       Yes
        & \cellcolor{red!5}
        No
        & \cellcolor{red!5}
        Yes  & \cellcolor{red!5} No  & \cellcolor{red!5} No 
        \\ 
    \hthickline
    \multirow{1}{*}{\rotatebox{0}{Sensor representation\ }} 
        & \cellcolor{blue!5}
        Single state
        & \cellcolor{blue!5}
       Single state
        & \cellcolor{blue!5}
        Trajectory
        & \cellcolor{blue!5}
        Trajectory  & \cellcolor{blue!5} Single State  & \cellcolor{blue!5} Trajectory 
        \\ 
    \hthickline
    \multirow{1}{*}{\rotatebox{0}{Processing \ }}    & \cellcolor{red!5}
        Snapshot
        & \cellcolor{red!5}
       Filter-based
        & \cellcolor{red!5}
       Filter-based
        & \cellcolor{red!5}
        Batch processing & \cellcolor{red!5} Filter-based & \cellcolor{red!5} Filter-based
        \\
    \hthickline
    \multirow{1}{*}{\rotatebox{0}{Time evolution\ }} 
        & \cellcolor{blue!5}
        No
        & \cellcolor{blue!5}
       Yes
        & \cellcolor{blue!5}
        Yes
        & \cellcolor{blue!5}
        Yes  & \cellcolor{blue!5} Yes  & \cellcolor{blue!5} Yes 
        \\ 
    \hthickline
     \multirow{1}{*}{\rotatebox{0}{SLAM Algorithm \ }}    & \cellcolor{red!5}
        MLE
        & \cellcolor{red!5}
       EKF
        & \cellcolor{red!5}
       RBPF + EKFs
        & \cellcolor{red!5}
       MLE & \cellcolor{red!5} BP & \cellcolor{red!5} RBPF + RFSs 
        \\
    \hthickline

    \end{tabular}
    \vspace{-5mm}
\label{tab:summary_of_differentSLAM}
\end{table*}

\vspace{-4mm}
\subsection{Models for Radio SLAM}

The \ac{SLAM} problem typically encompasses three main tasks: estimating the sensor's trajectory, constructing a map of the surrounding environment, and evaluating the associated uncertainties \cite{durrant2006simultaneous}. In probabilistic form, the \ac{SLAM} problem requires determining or approximating the joint posterior distribution of the sensor's trajectory and the environmental map. Focusing specifically on radio \ac{SLAM}, which leverages radio signals, the \ac{UE} functions as a sensor with an unknown and time-varying state. Static objects within the environment serve as landmarks, collectively forming the map \cite{ge2022mmwave}.

In terms of modeling, the fundamental challenge in radio SLAM is selecting \textit{suitable parametric representations for the UE and landmark states}, which describe the problem with sufficient accuracy. Typically, the \ac{UE} state is represented by the position, orientation, and clock parameters (due to unsynchronized \ac{BS} and \ac{UE}), whereas the landmark state is represented by the position \cite{ge2022mmwave}. Augmenting the \ac{UE} state with, for example, velocity and acceleration components or the landmark state with elements describing the spatial extent (i.e. shape, size, and orientation) or other characteristics like roughness is possible. However, there is a trade-off since every additional state element increases the system complexity and computational overhead of the algorithm. The \textit{state transition model} describes how the system state evolves over time, whereas the \textit{measurement model} describes how the channel parameters depend on the \ac{UE} and landmark state. In general, the probabilistic state transition and measurement models as well as the joint posterior density are represented using parametric distributions (e.g., multivariate Gaussian distributions), since they make it possible to approximate the Bayesian filtering equations in closed-form \cite{durrant2006simultaneous}. The joint posterior density can also be represented by a set of weighted particles that allow Monte Carlo approximations to form the solutions of the Bayesian filtering equations \cite{durrant2006simultaneous}.

In radio \ac{SLAM}, common transition models for the \ac{UE} include the constant-velocity model and coordinated turn model, whereas the landmarks are either assumed static or evolve according to a constant-velocity model. The measurement model is significantly more complicated and deserves a more comprehensive treatment. Ideally, the measurement model captures the type of propagation mechanism as well as the complex interaction of radio signals with the environment. A widely adopted approach is to use a multiple model method in which each model describes one propagation mechanism and interaction type \cite{ge2022mmwave}. For example, one model describes the direct \ac{LoS} transmission from the \ac{BS} to the \ac{UE}, a second model describes single-bounce reflections, and a third model describes single-bounce scattering. In radio \ac{SLAM}, the propagation mechanism is indirectly estimated using observations at different \ac{UE} locations. Referring to Fig.~\ref{fig:Radio SLAM overview}, if the \acp{IP} remain nearly constant as the \ac{UE} moves,  the environmental landmark is considered to be a small scattering object. If the \acp{IP} move deterministically together with the \ac{UE} movement, the environmental landmark is considered a reflecting surface. Reflecting surfaces are typically parameterized by a fixed \ac{VA} which is obtained by mirroring the \ac{BS} with respect to the surface so that both landmark types can be parameterized by a 3D location. 
The features of the communication system and the channel parameter estimator can also be incorporated into the probabilistic measurement model by modeling the clutter and probability of detection. The clutter model describes false detections of the channel parameter estimator or the presence of noise peaks unrelated to physical propagation paths of interest, leading to numerous false measurements. Whereas the latter describes the probability of detecting a landmark with the given communication system parameters, as a result, some landmarks within the \ac{FoV} may not be identified by the \ac{UE}.

\vspace{-4mm}
\subsection{Radio SLAM Methods}
The radio \ac{SLAM} problem is inherently complex due to multiple factors \cite{ge2022mmwave}. The primary difficulty arises from the uncertain number of landmarks, as the \ac{UE} does not have prior information on how many landmarks exist within its \ac{FoV} because the map is unknown. Additionally, clutter measurements pose a significant problem, increasing the risk of incorrect detections. Another obstacle is the imperfect detection capability, leading to incomplete or inaccurate mapping. Furthermore, there is the issue of unknown \ac{DA}, as the \ac{UE}  lacks information about the origin of each measurement. This creates a fundamental challenge in determining whether a measurement corresponds to an already detected landmark, a new landmark, or is simply clutter resulting in outliers that need to be removed. Moreover, another challenge is to deal with the correlation between the sensor state UE and the map state. Keeping track of the cross-correlation between the UE trajectory and the map will preserve all information but usually requires high complexity. However, throwing out this cross-correlation will constitute an inherent loss of information and is the price to pay for reducing complexity. Effective radio \ac{SLAM} solutions must address all these challenges comprehensively while maintaining manageable levels of signal processing complexity.

Radio \ac{SLAM} can be solved by various classes of methods \cite{ge2022mmwave}. At one end of the spectrum are the \textit{snapshot SLAM methods}, which treat each time step separately, and solve the SLAM problem at each time step only using all the measurements from that time step without considering any other measurements or mapping and positioning information from other time steps. These methods basically perform an optimization problem on the sensor state at the current time step and the map given all measurements from that time step. At the other end of the spectrum are the \textit{batch-based methods}, which process a sequence of measurements. In between are the recursive, \textit{filter-based methods}, which rely on Bayesian filtering.  
The \textit{\ac{EK}-SLAM} utilizes the Gaussian property of the SLAM posterior and employs the \ac{EKF} to estimate the posterior, where the approximation of the nonlinearity of the models is formed by using the first-order Taylor series. However, EK-SLAM linearizes the problem for efficiency, but struggles with highly nonlinear models. In addition, correlations between the UE state and the map can sometimes be dropped. \textit{FastSLAM}, based on the conditional independence of the landmarks given the UE trajectory, factorizes the SLAM posterior into separate distributions for the landmarks conditioned on the UE trajectory and the UE trajectory itself. Following the \ac{RBPF}, FastSLAM uses multiple weighted particles to represent the UE trajectory, each maintaining its own map, and SLAM posterior is then propagated through these particles and conditional maps. However, this approach requires a large number of particles to perform effectively, resulting in high computational complexity.  Both FastSLAM and EK-SLAM require pre-resolving DA, making them vulnerable to errors in DA. \textit{GraphSLAM} also addresses the DA challenge in advance and then represents the SLAM problem as a graphical model, where nodes represent the UE states and landmarks, and the edges represent their relationships. This batch processing approach opts for batch estimation, which usually works offline and can produce more accurate and robust estimates but relies heavily on reliable DA. \textit{\Ac{BP}-SLAM} offers a unified framework that incorporates DA within the estimation process by introducing auxiliary variables and using belief propagation on a factor graph. While BP-SLAM can handle complex scenarios without pre-resolving DA, it cannot explicitly track correlations between the UE state and the map and often requires additional adjustments for dynamic landmarks. \textit{\Ac{RFS}-SLAM}  provides another effective approach by modeling landmarks and measurements as sets with a random number of elements, which naturally handles uncertainties in the number and states of landmarks and DA uncertainties. Computing the joint posterior of the UE trajectory and the map utilizing RFS statistics, offering flexibility and robustness in handling dynamic and uncertain environments. A summary of the algorithms mentioned above is in Table~\ref{tab:summary_of_differentSLAM}.

 \begin{figure*}
\centering
\includegraphics[width=0.8\linewidth]{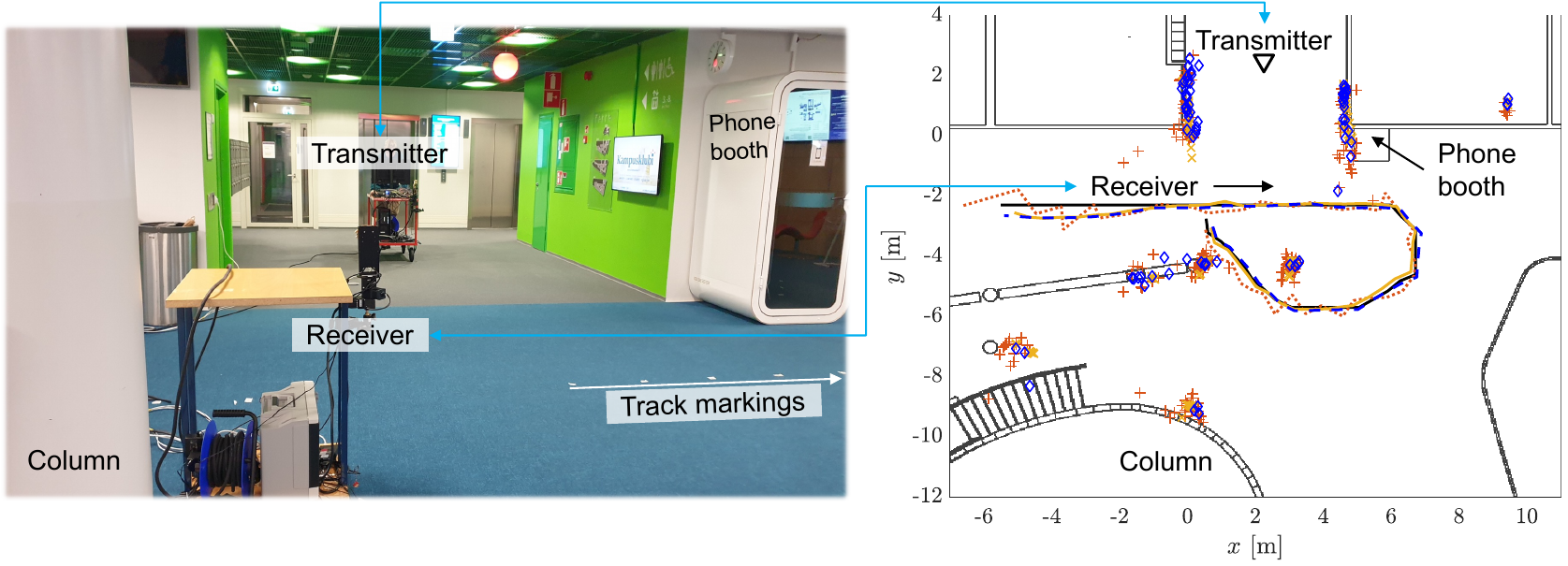}
		\vspace{-4mm}
        \caption{Performance comparison of snapshot SLAM (in red), RFS-SLAM (in yellow), and GraphSLAM (in blue). The root mean-squared errors (RMSEs) for position, heading, and clock bias across multiple Monte Carlo simulations are 0.35~m, 2.05\textdegree, and 1.42~ns for snapshot algorithm; 0.33~m, 2.04\textdegree, and 0.94~ns for RFS-SLAM; and 0.26~m, 1.78\textdegree, and 0.56~ns for GraphSLAM. GraphSLAM demonstrates the best mapping accuracy, with landmark estimates closely aligned with the floorplan, whereas snapshot SLAM shows the poorest performance, characterized by several highly inaccurate landmark estimates. The figure presents a single realization of the results. \rev{The results demonstrate that all three methods can estimate the UE trajectory and map the surrounding environment, despite the lack of synchronization between the BS and UE.}}   
        \vspace{-3mm}
        \label{fig:slam_comparison}
\end{figure*}

 \vspace{-5mm}
\subsection{Towards the Standardization of SLAM Back-End}

Standardization of radio SLAM is beneficial for interoperability in various applications, including smart cities, ADAS, UAVs, and industrial automation \rev{\cite{kaushik2024toward}}. These efforts aim to optimize the telecommunication infrastructure for positioning, sensing, and SLAM services while ensuring accuracy, reliability, and efficiency.
Key standardization areas include signaling protocols for BS and landmark locations, \rev{and service procedures}. The defined KPIs address positioning accuracy, velocity estimation, latency, refresh rates, and false alarm rates, ensuring robust performance in use cases such as collision avoidance and intrusion detection. SLAM data integration across 3GPP and non-3GPP sensors (e.g., radar, cameras) further enhances system adaptability. Privacy and security are critical, requiring strict access control, encryption, and regulatory compliance to prevent unauthorized access or data interception. Balancing standardization with flexibility allows for vendor-driven optimizations while maintaining common data formats for cooperative SLAM and sensor fusion.

\vspace{-5mm}
\subsection{Radio SLAM Front-End  -- a Case Study}

To compare radio SLAM algorithms, we evaluate snapshot SLAM, RFS-SLAM, and GraphSLAM using experimental data  (see Fig.~\ref{fig:slam_comparison})\rev{, where the BS and UE are unsynchronized in a bi-static scenario, and SLAM is performed at the UE side}. RFS-SLAM, a representative filter-based method, is selected for its robustness in dynamic environments (Table~\ref{tab:summary_of_differentSLAM}). The 60 GHz \rev{(part of the 5G NR band n263, selected for its unlicensed nature)} experiments, conducted at 45 UE locations, used 400 MHz bandwidth and \rev{beamformed} 5G NR downlink positioning signals\rev{, where the \ac{EIRP} of the transmitted signals varies from \(33\) to \(36\)~dBm, depending on the beamforming angles used.} 
Further experimental details are in \cite{rastorgueva2024millimeter}. All three algorithms perform well, but snapshot SLAM is the weakest, as it ignores temporal correlations and relies solely on current measurements. RFS-SLAM improves upon this by incorporating a Bayesian filtering approach that leverages motion modeling. GraphSLAM outperforms RFS-SLAM by employing batch processing, which optimally integrates past and future measurements. In contrast, filter-based methods (e.g., RFS-SLAM) process data sequentially, relying only on measurements up to the current time step.

\vspace{-4mm}
\section{Outlook}

The use of radio SLAM in future communication systems presents new opportunities and challenges, which are described below and visualized in Fig.~\ref{fig:Opportunities and challenges}.

\begin{itemize}
    \item 
\textit{Complementary sensing:}
Radio SLAM complements traditional radar sensing and GNSS-based positioning,  providing redundancy at low cost.  
\item \textit{Cooperative SLAM:}
Dense 6G networks with many BSs and UEs enable short-term cooperation via inter-BS and inter-UE sensing and long-term cooperation through map exchange and location sharing, \rev{all subject to privacy and data reliability considerations}.
\item \textit{Semantic SLAM:}
Integrating external sensors (e.g., cameras, radar) and AI-driven approaches 
can enhance positioning and sensing,  enable high-fidelity 3D mapping, capturing object sizes, materials, and defects, supporting applications like digital twins. 

\item \textit{Multiband SLAM:} 
Using multiple frequency bands simultaneously provides a diverse perspective on landmarks, enabling both through-the-wall and off-the-wall sensing.

\item \textit{Multi-bounce SLAM:}
Multi-bounce reflections, traditionally seen as interference, can be leveraged to detect occluded landmarks and improve mapping.

\begin{figure}
    \centering
\includegraphics[width=0.8\linewidth]{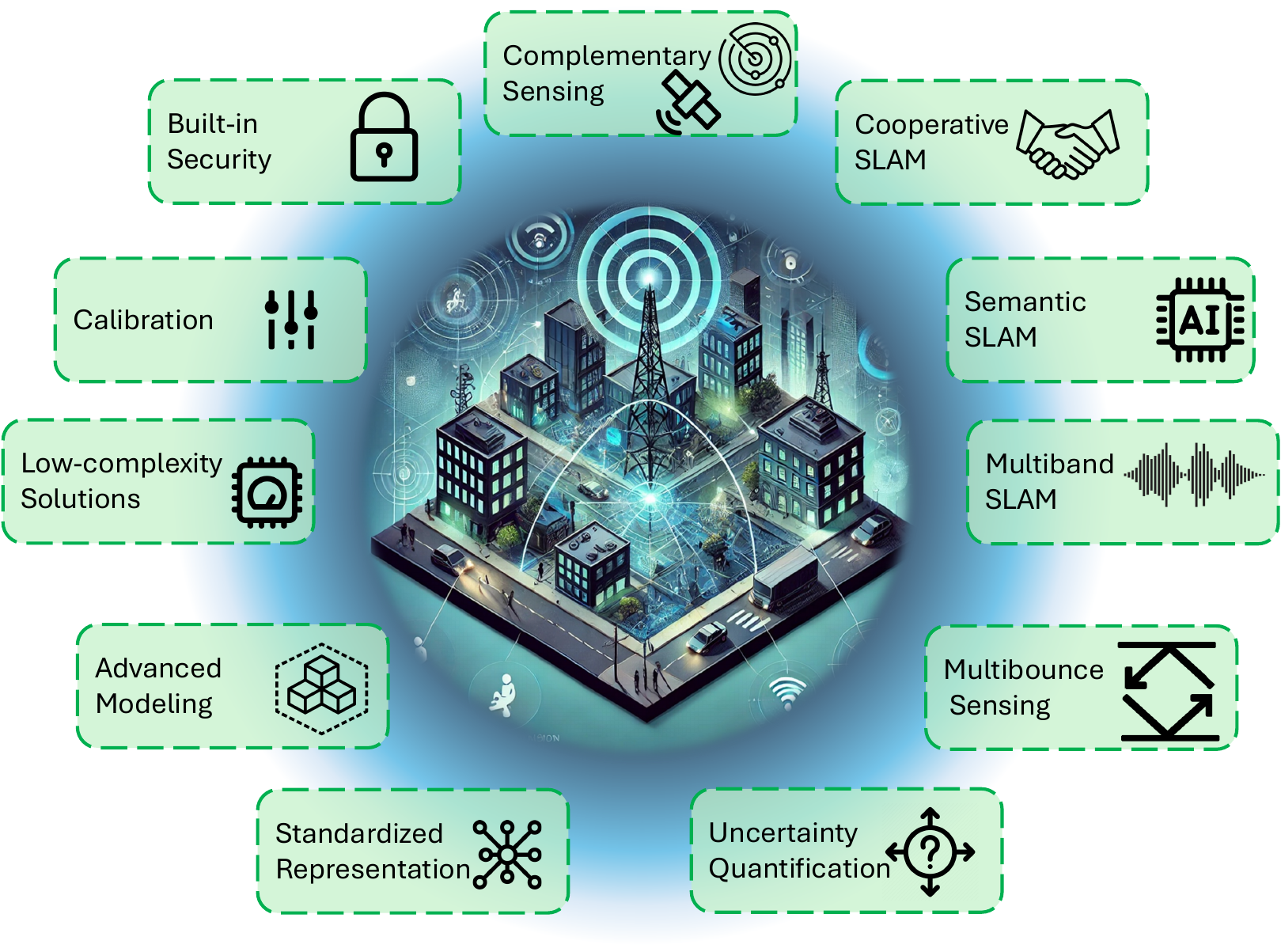}
\vspace{-3mm}
    \caption{Opportunities and challenges for 6G radio SLAM.} \vspace{-5mm} 
    \label{fig:Opportunities and challenges}
\end{figure}

\item \textit{Uncertainty quantification:}
Accurate uncertainty modeling of users and landmarks is needed for SLAM integrity and reliability, especially for safety-critical applications.

\item \textit{Standardization:}
Pilot signals measurement protocols must be defined,  SLAM must be integrated within the 6G network architectures (including both the location management function and the sensing management function). 
\item \textit{Advanced modeling:}
Accurate mobility models are needed for both connected users, static and moving objects. \rev{SLAM in near-field scenarios requires accurate channel modeling, necessitating corresponding adaptations in channel estimation algorithms.}

\item  \textit{Low-complexity solutions:}
Real-time SLAM is computationally demanding due to complex interactions, landmark diversity, and the need for concurrent hypotheses. Reducing complexity while maintaining high accuracy is crucial. Since most environments are relatively static, re-mapping can be minimized to optimize resources.
\item \textit{Calibration:}
BS and UE locations and orientations must be accurately calibrated to prevent positioning errors. Environmental factors further complicate calibration.

\item \textit{Security:}
Radio SLAM relies on shared sensing and positioning data, raising concerns about privacy, spoofing, and unauthorized access. Robust encryption, secure data sharing, and privacy-preserving techniques are essential to protect users and infrastructure.

\end{itemize}

\vspace{-4mm}
\section{Conclusions}
This paper provides an overview of radio SLAM as a key enabler of 6G ISAC, analyzing both front- and back-end aspects, including channel modeling, parameter estimation, and algorithmic approaches. We showcase radio SLAM performance across different frequency bands, highlighting the trade-offs between coverage, resolution, and computational complexity. 
Despite significant progress in radio SLAM, many  challenges remain, including multi-bounce exploitation,  uncertainty quantification, standardization, and calibration. We outline emerging opportunities, such as cooperative SLAM, multiband SLAM, and AI-enhanced modeling, which could enhance situational awareness and mapping accuracy. 
In our view, future research should focus on developing energy-efficient and low-complexity radio SLAM solutions for real-time applications; 
enhancing privacy-preserving mechanisms to protect sensitive localization and mapping data; refining AI-driven algorithms for improved adaptability in dynamic environments; and standardizing key SLAM components to ensure seamless integration in 6G.

\vspace{-5mm}
\section*{Acknowledgment}
This work was supported by the Swedish Research Council (VR) through the projects 6G-PERCEF (Grant 2024-04390) and HAILS (Grant 2022-03007), by the SNS JU project 6G-DISAC under the EU's Horizon Europe research and innovation Program under Grant Agreement No.~101139130, by the Research Council of Finland under the grants \#352754, \#357730, and \#359095, and by Business Finland through the 6G-ISAC project.

\vspace{-4mm}

\bibliography{IEEEabrv,Bibliography}
\vspace{-3mm}
\section*{Biographies}

\vskip -2.7\baselineskip plus -1fil

\begin{IEEEbiographynophoto}
{Yu Ge} is a Postdoctoral Researcher with the Department of Electrical Engineering at Chalmers University of Technology, Sweden.
\end{IEEEbiographynophoto}

\vskip -2.7\baselineskip plus -1fil

\begin{IEEEbiographynophoto}
{Ossi Kaltiokallio} is a Senior Research Fellow with the Electrical Engineering Unit at Tampere University, Finland. 
\end{IEEEbiographynophoto}

\vskip -2.7\baselineskip plus -1fil

\begin{IEEEbiographynophoto}
{Elizaveta Rastorgueva-Foi} is a doctoral candidate at the Unit of Electrical Engineering at Tampere University, Finland.
\end{IEEEbiographynophoto}

\vskip -2.7\baselineskip plus -1fil

\begin{IEEEbiographynophoto}
{Musa Furkan Keskin} is a Research Specialist with the Department of Electrical Engineering at Chalmers University of Technology, Sweden.
\end{IEEEbiographynophoto}

\vskip -2.7\baselineskip plus -1fil

\begin{IEEEbiographynophoto}
{Hui Chen} is a Research Specialist with the Department of Electrical Engineering at Chalmers University of Technology, Sweden.
\end{IEEEbiographynophoto}

\vskip -2.7\baselineskip plus -1fil

\begin{IEEEbiographynophoto}
{Guillaume Jornod} is a Researcher Engineer and 3GPP Delegate at Bosch, Germany.
\end{IEEEbiographynophoto}

\vskip -2.7\baselineskip plus -1fil

\begin{IEEEbiographynophoto}
{Jukka Talvitie} is a University Lecturer with the Department of Electrical Engineering at Tampere University, Finland. 
\end{IEEEbiographynophoto}

\vskip -2.7\baselineskip plus -1fil

\begin{IEEEbiographynophoto}
{Mikko Valkama} is a Professor and Department Head of Electrical Engineering at Tampere University, Finland. 
\end{IEEEbiographynophoto}

\vskip -2.7\baselineskip plus -1fil

\begin{IEEEbiographynophoto}
{Frank Hofmann} is the Chief Expert and Group Leader in the field of communication systems at Bosch, Germany.
\end{IEEEbiographynophoto}

\vskip -2.7\baselineskip plus -1fil

\begin{IEEEbiographynophoto}
{Henk Wymeersch} is a Professor with the Department of Electrical Engineering at Chalmers University of Technology, Sweden.
\end{IEEEbiographynophoto}

\end{document}

%% file: acronyms.tex
\acrodef{RIS}{reconfigurable intelligent surface}
\acrodef{SNR}{signal-to-noise ratio}
\acrodef{ISAC}{integrated sensing and communication}
\acrodef{ISLAC}{integrated sensing, localization, and communication}
\acrodef{SLAM}{simultaneous localization and mapping}
\acrodef{LoS}{line-of-sight}
\acrodef{NLoS}{non-line-of-sight}
\acrodef{AoA}{angle-of-arrival}
\acrodef{AoD}{angle-of-departure}
\acrodef{ToA}{time-of-arrival}
\acrodef{UE}{user equipment}
\acrodef{NF}{noise figure}
\acrodef{PSD}{power spectral density}
\acrodef{BS}{base station}
\acrodef{MCRB}{misspecified Cram\'{e}r-Rao bound}
\acrodef{CRB}{Cram\'{e}r-Rao bound}
\acrodef{LB}{lower bound}
\acrodef{ML}{maximum-likelihood}
\acrodef{MML}{mismatched maximum-likelihood}
\acrodef{DL}{downlink}
\acrodef{UL}{uplink}
\acrodef{FF}{far-field}
\acrodef{MIMO}{multiple-input multiple-output}
\acrodef{MISO}{multiple-input single-output}
\acrodef{SISO}{single-input single-output}
\acrodef{SIP}{shift invariance property}
\acrodef{FIM}{Fisher information matrix}
\acrodef{RMSE}{root mean-squared error}
\acrodef{RMS}{root mean-squared}
\acrodef{AWGN}{additive white Gaussian noise}
\acrodef{ADMM}{alternating direction method of multipliers}
\acrodef{LS}{least-squares}
\acrodef{SOC}{second-order cone}
\acrodef{CFO}{carrier frequency offset}
\acrodef{GLRT}{generalized likelihood ratio test}
\acrodef{FSPL}{free space path loss}
\acrodef{TDoA}{time-difference-of-arrival}
\acrodef{MPC}{multi-path components}
\acrodef{NB}{narrowband}
\acrodef{WB}{wideband}
\acrodef{TDM}{time-division multiplexing}
\acrodef{NTN}{non-terrestrial network}
\acrodef{KPI}{key performance indicator}
\acrodef{KVI}{key value indicator}
\acrodef{SDG}{sustainable development goal}
\acrodef{LEO}{low earth orbit}
\acrodef{OFDM}{orthogonal frequency division multiplexing}
\acrodef{LCS}{local coordinate system}
\acrodef{GCS}{global coordinate system}
\acrodef{CP}{cyclic prefix}
\acrodef{BF}{beamforming}
\acrodef{CRLB}{Cramér-Rao lower bound}
\acrodef{MLE}{maximum-likelihood estimator}
\acrodef{URA}{uniform rectangular array}
\acrodef{FDoA}{frequency-difference-of-arrival}
\acrodef{UAV}{unmanned aerial vehicle}
\acrodef{DA}{data association}
\acrodef{FoV}{field of view}
\acrodef{EKF}{extended Kalman filter}
\acrodef{EK}{extended Kalman}
\acrodef{RBPF}{Rao-Blackwellized particle filter}
\acrodef{BP}{belief propagation }
\acrodef{RFS}{random finite set }
\acrodef{PPP}{Poisson point process}
\acrodef{IP}{interaction point}
\acrodef{VA}{virtual anchor}

\acrodef{ADAS}{advanced driving assistance system}
\acrodef{DFT}{discrete Fourier transform }

\acrodef{ESPRIT}{estimation of signal parameters via rotational invariant techniques}

\acrodef{MUSIC}{multiple signal classification}

\acrodef{SAGE}{space-alternating generalized expectation-maximization}
\acrodef{RiMAX}{a maximum likelihood parameter estimation framework}

\acrodef{CDF}{cumulative distribution function}
\acrodef{EIRP}{effective isotropic radiated power}